\documentclass[conference]{IEEEtran}
\IEEEoverridecommandlockouts
\usepackage{cite}
\usepackage{amsmath,amssymb,amsfonts}
\usepackage{algorithmic}
\usepackage{graphicx}
\usepackage{textcomp}
\usepackage{booktabs}
 \usepackage[nolist,nohyperlinks]{acronym}
\usepackage{xcolor}
\def\BibTeX{{\rm B\kern-.05em{\sc i\kern-.025em b}\kern-.08em
    T\kern-.1667em\lower.7ex\hbox{E}\kern-.125emX}}

\usepackage[acronym]{glossaries}
\acrodef{mIoU}{mean intersection over union}
\acrodef{DSC}{dice coefficient}
\IEEEoverridecommandlockouts
\IEEEpubid{\makebox[\columnwidth]{978-1-6654-0810-3/21/\$31.00~\copyright~2021~IEEE \hfill}
\hspace{\columnsep}\makebox[\columnwidth]{ }}
\begin{document}

\title{PAANet: Progressive Alternating Attention for Automatic Medical Image Segmentation\\}

\author{\IEEEauthorblockN{Abhishek Srivastava\IEEEauthorrefmark{1}, Sukalpa Chanda\IEEEauthorrefmark{2}, Debesh Jha\IEEEauthorrefmark{3}\IEEEauthorrefmark{4}, Michael A. Riegler\IEEEauthorrefmark{3}\IEEEauthorrefmark{4},\\ P{\aa}l Halvorsen\IEEEauthorrefmark{3}\IEEEauthorrefmark{5}, Dag Johansen\IEEEauthorrefmark{4}, Umapada Pal\IEEEauthorrefmark{1}
}
\vspace{2mm}

\IEEEauthorblockA{\IEEEauthorrefmark{1}Indian Statistical Institute, Kolkata, India\ \ \ \ \
\IEEEauthorrefmark{2}Østfold University College, Norway\ \ \ \ \ \
\IEEEauthorrefmark{3}SimulaMet, Norway \\
\IEEEauthorrefmark{4}UiT The Arctic University of Norway, Norway \ \ \ \ \ \
\IEEEauthorrefmark{5}Oslo Metropolitan University, Norway\\
Email: {abhisheksrivastava2397@gmail.com}}}

\maketitle
\begin{abstract}
Medical image segmentation can provide detailed information for clinical analysis which can be useful for scenarios where the detailed location of a finding is important. Knowing the location of a disease can play a vital role in treatment and decision-making. Convolutional neural network (CNN) based encoder-decoder techniques have advanced the performance of automated medical image segmentation systems. Several such CNN-based methodologies utilize techniques such as spatial- and channel-wise attention to enhance performance. Another technique that has drawn attention in recent years is residual dense blocks (RDBs). The successive convolutional layers in densely connected blocks are capable of extracting diverse features with varied receptive fields and thus, enhancing performance. However, consecutive stacked convolutional operators may not necessarily generate features that facilitate the identification of the target structures. In this paper, we propose a progressive alternating attention network (PAANet). We develop progressive alternating attention dense (PAAD) blocks, which construct a guiding attention map (GAM) after every convolutional layer in the dense blocks using features from all scales. The GAM allows the following layers in the dense blocks to focus on the spatial locations relevant to the target region. Every alternate PAAD block inverts the GAM to generate a reverse attention map which guides ensuing layers to extract boundary and edge-related information, refining the segmentation process. Our experiments on three different biomedical image segmentation datasets exhibit that our PAANet achieves favorable performance when compared to other state-of-the-art methods.
\end{abstract}


\section{Introduction}
Medical image segmentation based on deep learning methods have garnered a lot of attention during the past few years. In this context, to generate segmentation datasets, manual annotation of images to identify and locate the region-of-interest is a time consuming task. The accuracy of such annotation depends upon the expertise of the medical professionals making it prone to undesired oversights. Convolutional Neural Network (CNN) based techniques for delineation of desired anatomical regions from medical images have been proven to be effective. Methods like U-Net~\cite{ronneberger2015u}, U-Net++~\cite{zhou2019unet++}, ResUNet++~\cite{jha2019resunet++}, DoubleUNet~\cite{jha2020doubleu} PraNet~\cite{fan2020pranet}, and Attention U-Net~\cite{oktay2018attention} have served as baseline approaches achieving good segmentation performance. Attention mechanisms have served as an integral component in such methods. Attention U-Net uses the deep features from lower levels of the decoder to generate spatial attention maps, which are in turn used to prune irrelevant features from skip-connections. ResUNet++~\cite{jha2019resunet++} used the squeeze and excitation block~\cite{hu2018squeeze} to model inter-dependencies between the channels to suppress irrelevant and enhance relevant channels. FED-Net~\cite{chen2019feature} introduced feature fusion blocks which applied a combination of spatial- and channel-wise attention to increase the network's segmentation ability. Different combinations of channel and spatial attention at various stages of the encoder-decoder structure have also been used in FocusNet~\cite{kaul2019focusnet} and MSRF-Net~\cite{srivastava2021msrf}. PNS-Net~\cite{ji2021progressively} designed a self-attention mechanism for video polyp segmentation to utilize both temporal and spatial features. PraNet~\cite{fan2020pranet} used reverse attention by generating an initial global guiding map which was later used for mining boundary cues. Another familiar technique for image segmentation in both medical imaging and natural computer vision is residual dense blocks (RDBs)~\cite{zhang2018residual,zhang2018road,dolz2018hyperdense,ibtehaz2020multiresunet}. The main advantage offered by RDBs is the combination of features obtained by both high- and low-receptive fields. These advantages motivated many works to incorporate RDBs in encoder-decoder based architectures~\cite{yang2019road,ding2019deep,srivastava2020residual}. An important aspect of RDBs is that multiple convolutional layers with a smaller number of output channels are stacked on top of each other. This allows to progressively increase the receptive field while maintaining relevant low-level features. However, the feature maps produced by such consecutive convolutional layers rely on the gradients received by the successive decoder layers for capturing the region of interest. Providing additional intermediate supervision for each convolutional layer within an RDB may help in generation of meaningful and informative features.

\begin{figure*}[!t]
    \centering
    \includegraphics[width=0.95\textwidth]{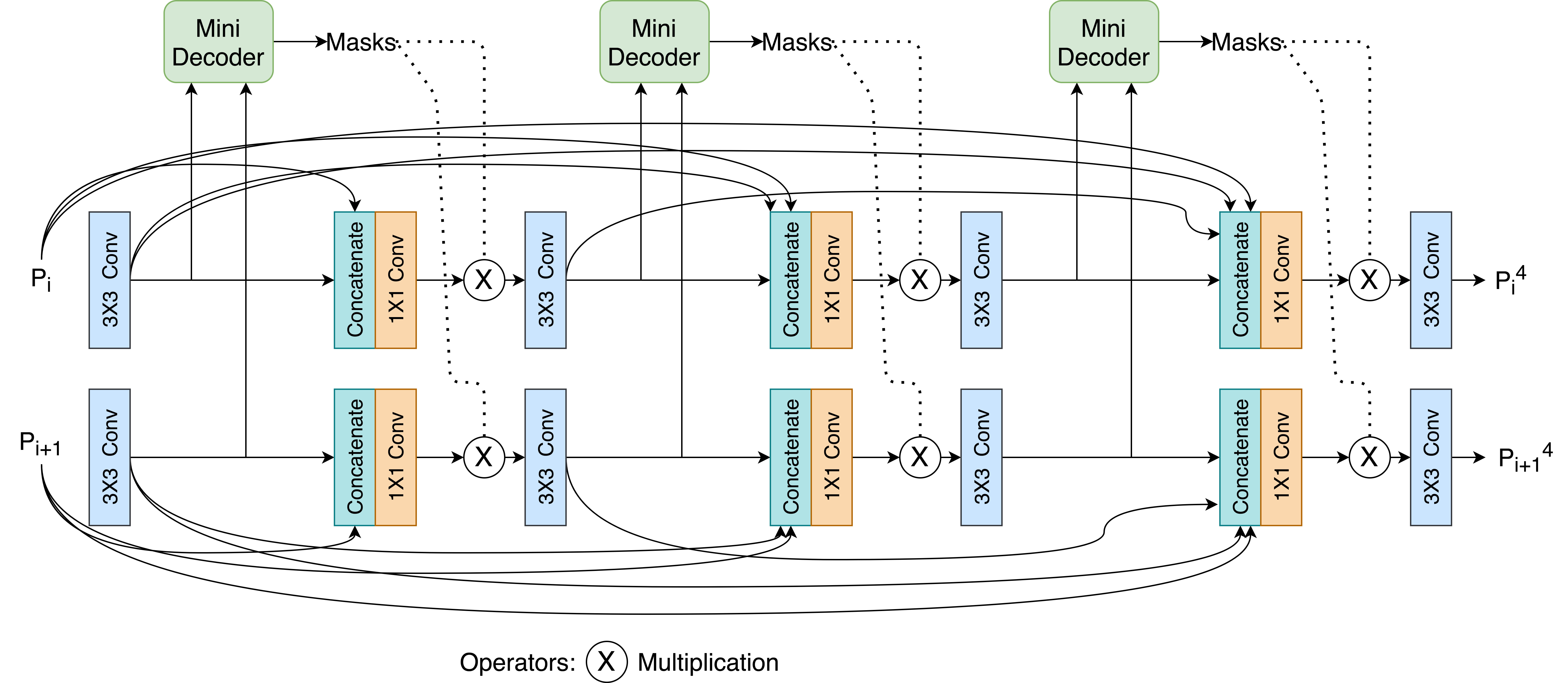}
    \caption{The architecture of our Progressive Alternative Attention Dense (PAAD) block}
    \label{fig:PAANet}
\end{figure*}

\begin{figure}[!t]
    \centering
    \includegraphics[width=\linewidth]{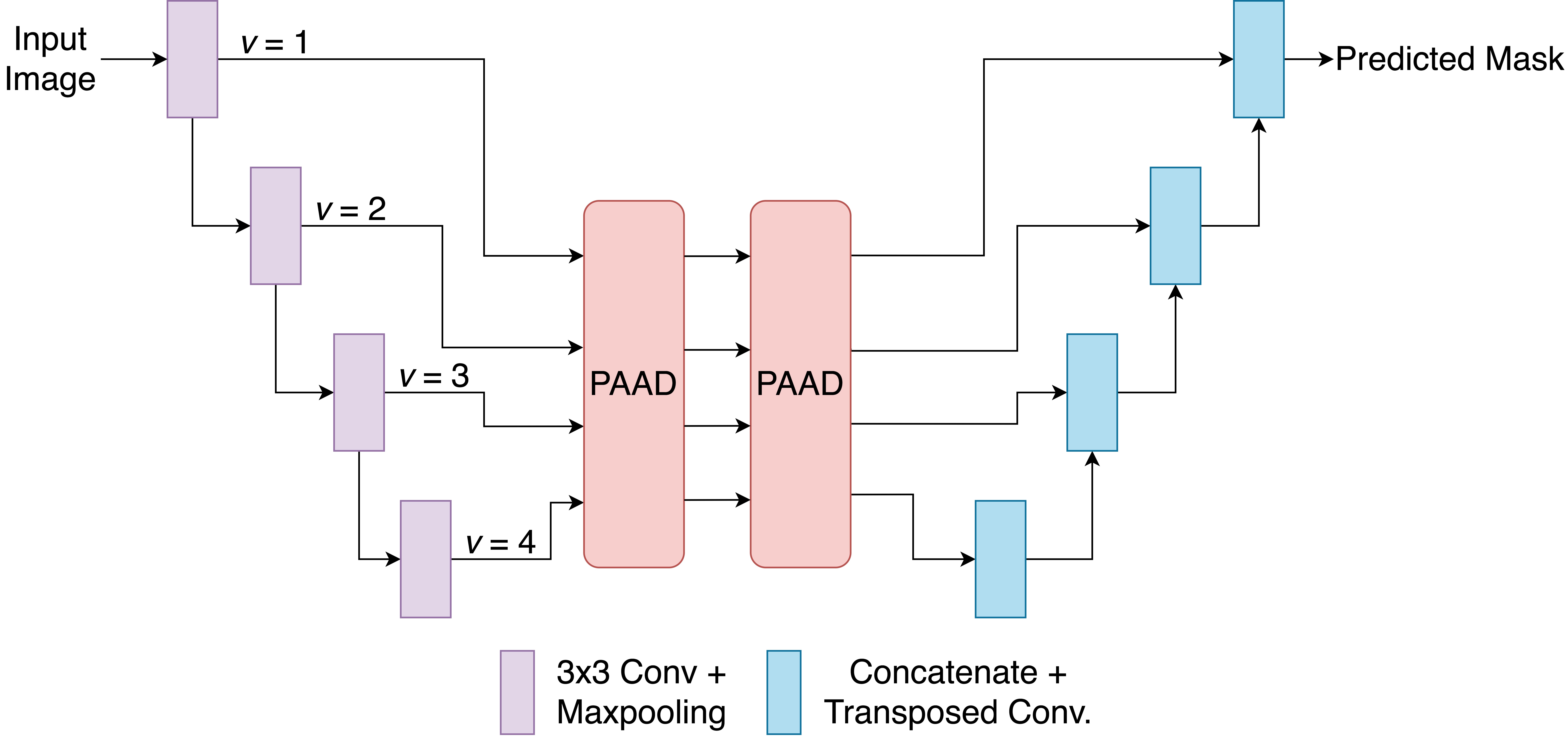}
    \caption{Overview of the complete PAANet architecture.}
    \label{fig:PAAD_FULL}
\end{figure}

In this work, we introduce a novel progressive alternating attention dense block (PAAD). After each convolutional layer in the dense block, a mini-decoder is used to generate a guiding attention map (GAM). The successive layers utilize this segmentation map to prune features impertinent to identifying the region of interest. Additionally, we use reverse attention in every alternative PAAD block which allows layers to further mine peripheral features allowing the network to accurately capture the variation in shape and size of the region of interest. Since the GAM is created after every convolutional layer of the PAAD blocks, they are updated progressively which further refine the quality of feature maps, prune irrelevant features, and allow the later convolutional layers to produce only meaningful features. We validate our method on three different biomedical datasets: Data Science Bowl (DSB) 2018~\cite{caicedo2019nucleus}, ISIC 2018 skin lesion segmentation~\cite{codella2018skin,tschandl2018ham10000}, and Kvasir-Instruments~\cite{jha2021kvasir}.

\section{Method}
In this section, we elaborate upon the encoder PAAD block used in our PAANet. The architecture of our progressive alternating attention network (PAANet) is illustrated in Figure~\ref{fig:PAAD_FULL}. The input image is encoded using ResNet-50 pretrained on ImageNet~\cite{deng2009imagenet}, which has served as a standard backbone for medical image segmentation models. Let all features from different levels of the encoder be denoted as $E_{v}$, where $v \in {1,2,3,4}$.

\subsection{Progressive Alternating Attention Dense Blocks}
Each feature set from the encoder blocks is fed into the PAAD blocks. Equation~\ref{eq:dense} describes how the feature maps are generated for each convolutional layer within PAAD blocks:

\begin{equation}
    F_{v}^{c} = Con(P_{v}^{c-1} \oplus P_{v}^{c-2} \cdots \oplus P_{v}^{0})
    \label{eq:dense}
\end{equation}
Here, $F_{v}^{c}$ denotes the features generated by the layer number $c$ for the $v'th$ resolution scale, $\oplus$ represents the concatenation operator, and $Con$ represents a $3\times3$ convolutional operator. $P_{v}^{0}$ is initially set to $E_{v}$.
The architecture of the PAAD block is shown in Figure~\ref{fig:PAANet}. For clarity, we have only selected two feature streams with distinct resolution in the figure to represent the operations and the feature flow within the PAAD block.

\subsubsection{Mini-Decoder}
We elaborate upon the mechanism of our Mini-Decoder block (see Figure~\ref{fig:PAANet}) in Equation~\ref{eq:mini}:
\begin{equation}
    G^{c} = \sigma(Con(F_{1}^{c}\oplus F_{2}^{c}\oplus F_{3}^{c}\oplus F_{4}^{c}))
    \label{eq:mini}
\end{equation}
For all scales, $F_{v}^{c}$ is upscaled to the size of the ground truth map. They are concatenated before being processed by a convolutional layer with kernel size $3\times3 $. Finally, a sigmoid function is used to transform the feature maps within the range of [0,1]. The guiding segmentation map is supervised using the ground truth. 

\subsubsection{Progressive Alternating Attention}
In this stage, the feature maps generated by the layer $c$ in the PAAD block are multiplied by the guidance map $G$ as described in Equation~\ref{eq:mult}:
\begin{equation}
    P_{v}^{c} = F_{v}^{c} \otimes G_{v}^{c}
    \label{eq:mult}
\end{equation}
The guidance maps is downscaled appropriately to the spatial dimensions of the $v$'th scale. In the first PAAD block, we use spatial attention to allow convolutional layers to progressively prune irrelevant features while focusing on the region of interest as deemed by the guidance maps. In the next PAAD block, the GAM is inverted as $G_{v}^{c} = 1 - G_{v}^{c}$, to allow all layers of the PAAD to extract further boundary and edge information which may have been omitted in the first PAAD block. This results in further refining the feature maps and restricting the flow of extraneous features. We maintain informative deep level and shallow level features throughout the process, consequently, improving accuracy of our proposed PAANet. Further, skip connections are added from the input to improve gradient flow.

\subsection{Decoder}
 The final output from the PAAD is denoted as $P_{v}$. The output from each decoder level is upscaled to the spatial resolution dimensions of the next decoder level as shown in Equation~\ref{eq:deco}. The skip-connections from the corresponding scale is concatenated to the upscaled features before being fused using a $3\times3$ convolutional layer.
\begin{equation}
    P_{v} = Con(Upscale(P_{v-1}) \otimes P_{v})
    \label{eq:deco}
\end{equation}
Here, $Upscale$ represents a strided $4\times4$ transposed convolution layer. Each $P_{v}$ where $v \in {2,3,4}$ is upscaled to the spatial resolution of the ground truth and deeply supervised. Combination of equally weighted intersection over union loss and binary-cross entropy loss is used for supervision

\section{Experiments and Results}

To conduct experiments and determine the effectiveness of our method, we choose three different biomedical image segmentation datasets as the use case. These three different datasets have different types of segmentation masks. At first, we experimented with the 2018 Data Science Bowl (DSB) Challenge dataset~\cite{caicedo2019nucleus} which contains 670 segmented nuclei images. The nuclei images found in DSB 2018 were captured under varying conditions like different cell sizes, magnification, and imaging modality. This variation within the distribution makes the segmentation of nuclei images a challenging problem.  Next, we experimented using the ISIC-2018 Challenge~\cite{codella2018skin,tschandl2018ham10000} dataset, which is a skin segmentation dataset. Skin lesion segmentation assists in melanoma detection, melanoma being the most severe form of skin cancer, warrants an automatic skin lesion segmentation system. Therefore, developing automated systems could be helpful in the clinic. Our third dataset used in the experiments was the Kvasir-Instruments~\cite{jha2021kvasir} dataset which is a diagnostic and therapeutic tool segmentation dataset in gastrointestinal endoscopy. Tool segmentation in gastrointestinal images allows tracking of instruments used during endoscopy and could assist robotic and non-robotic surgeries. Developing such an automated segmentation system might help in complex real-time surgeries inside the gastrointestinal tract. 

\begin{table}[!t]

\centering

\scriptsize
\caption{Results on the 2018 Data Science Bowl}
\label{tab:result3}
\begin{tabular}{@{}l|l|l|l|l@{}}
\toprule
\textbf{Method} & \textbf{DSC} & \textbf{mIoU}& \textbf{Recall} & \textbf{Precision}\\ 
\hline
\hline
U-Net~\cite{ronneberger2015u} & 0.9080 & 0.8314 & 0.9029 & 0.9130    \\ \hline
U-Net++~\cite{zhou2019unet++}  & 0.7705 & 0.5265 & 0.7159 & 0.6657 \\ \hline
ResUNet++~\cite{jha2019resunet++} & 0.9098 & 0.8370 & 0.9169 & 0.9057 \\ \hline
Deeplabv3+ (Xception)~\cite{chen2018encoder} & 0.8857 & 0.8367 & 0.9141 & 0.9081 \\\hline
Deeplabv3+ (Mobilenet)~\cite{chen2018encoder} & 0.8239 & 0.7402 & 0.8896 & 0.8151 \\ \hline
HRNetV2-W18-Smallv2~\cite{wang2020deep} & 0.8495 & 0.7585 & 0.8640 & 0.8398\\ \hline 
HRNetV2-W48~\cite{wang2020deep} & 0.8488 & 0.7588 & 0.8359 & 0.8913\\ \hline
ColonSegNet~\cite{jha2020real} & 0.9197 & 0.8466 & 0.9153 & \textbf{0.9312} \\ \hline
ResUNet++ + CRF~\cite{jha2021comprehensive}& 0.7806 & 0.7322 & 0.7534 & 0.6308 \\ \hline
PraNet~\cite{fan2020pranet}  &0.8751 & 0.7868 & 0.9182 & 0.8438 \\ \hline
MSRF-Net~\cite{srivastava2021msrf}  &0.9224 & 0.8534 & \textbf{0.9402} & 0.9022 \\ \hline 
PAANet(Ours) & \textbf{0.9244} & \textbf{0.8627} & 0.9319 & 0.9208 \\ \hline
\bottomrule
\end{tabular}
\label{tab:result1}
\vspace{-5mm}
\end{table}
\begin{table}[!t]

\centering
\scriptsize
\caption{Results on the ISIC-2018 skin lesion segmentation challenge}
\label{tab:result4}
\begin{tabular}{@{}l|l|l|l|l@{}}
\toprule
\textbf{Method} & \textbf{DSC}  & \textbf{mIoU}& \textbf{Recall} & \textbf{Precision}\\  \hline
\hline
U-Net~\cite{ronneberger2015u} & 0.8554 & 0.7847 & 0.8204 & \textbf{0.9474} \\ \hline
U-Net++~\cite{zhou2019unet++}  & 0.8094 & 0.7288 & 0.7866 & 0.9084\\ \hline
ResUNet++~\cite{jha2019resunet++} & 0.8557 & 0.8135 & 0.8801 & 0.8676 \\ \hline
Deeplabv3+ (Xception)~\cite{chen2018encoder} & 0.8772 & 0.8128 & 0.8681 & 0.9272 \\\hline
Deeplabv3+ (Mobilenet)~\cite{chen2018encoder} & 0.8781 & 0.8236 & 0.8830 & 0.9244 \\ \hline
HRNetV2-W18-Smallv2~\cite{wang2020deep} & 0.8561 & 0.7821 & 0.8556 & 0.8974  \\ \hline 
HRNetV2-W48~\cite{wang2020deep} & 0.8667 & 0.8109 & 0.8584 & 0.9155  \\ \hline
ResUNet++ + CRF~\cite{jha2021comprehensive}& 0.8688 & 0.8209 & 0.8826 & 0.8736 \\ \hline
MSRF-Net~\cite{srivastava2021msrf} & 0.8824 & \textbf{0.8373} & 0.8893 & 0.9348  \\ \hline   
PAANet (Ours) & \textbf{0.8912} & 0.8219 & \textbf{0.9019} & 0.9054 \\ \hline
\bottomrule

\end{tabular}
\label{tab:result2}
\vspace{-5mm}
\end{table}

\begin{table}[!t]

\centering

\scriptsize
\caption{Results on the Kvasir-Instruments}
\label{tab:result3}
\begin{tabular}{@{}l|l|l|l|l@{}}
\toprule
\textbf{Method} & \textbf{DSC} & \textbf{mIoU}& \textbf{Recall} & \textbf{Precision}\\ 
\hline
\hline
U-Net~\cite{ronneberger2015u} & 0.9158 & 0.8578 & 0.9487 & 0.8998    \\ \hline
U-Net++~\cite{zhou2019unet++}  & 0.8808 & 0.8453 & 0.8623 & 0.9173\\ \hline
HRNetV2-W18-Smallv2~\cite{wang2020deep} & 0.9272 & 0.8822 & 0.9244 & 0.9438\\ \hline 
HRNetV2-W48~\cite{wang2020deep} & 0.9306 & 0.8867 & 0.9294 & 0.9429  \\ \hline
Deeplabv3+ (Xception)~\cite{chen2018encoder} & 0.8998 & 0.8615 & 0.9012 & 0.9272 \\\hline
Deeplabv3+ (Mobilenet)~\cite{chen2018encoder} & 0.9079 & 0.8635 & 0.9075 & 0.9468 \\ \hline
ColonSegNet~\cite{jha2020real} & 0.9201 & 0.8820 & 0.9169 & 0.9317\\ \hline
MSRF-Net~\cite{srivastava2021msrf} & 0.9379 & 0.8990 & \textbf{0.9661} & 0.9283  \\ \hline
PAANet(Ours) & \textbf{0.9495} & \textbf{0.9160}  & 0.9475 & \textbf{0.9571} \\
\hline
\bottomrule
\end{tabular}
\label{tab:result3}
\vspace{-5mm}
\end{table}

In our experiment, we have used Adam optimizer with a learning rate of $1e-4$. All models were trained for 30 epochs with a batch size of 8. The DSB and ISIC-2018 datasets were divided into training, validation, and testing splits which contains 80\%, 10\%, and 10\% of the data, respectively. For Kvasir-Instruments, we follow the official train-test split provided by the dataset provider. We have compared our work with state-of-the-art (SOTA) medical image segmentation methods such as U-Net~\cite{ronneberger2015u},U-Net++~\cite{zhou2019unet++}, ``ResUNet++ + CRF~\cite{jha2021comprehensive}''. Further comparisons were made with standard semantic segmentation methods like HR-Net~\cite{wang2020deep} and Deeplabv3+~\cite{chen2018encoder}. 

\begin{figure}[!t]
    \centering
    \includegraphics[width=\linewidth]{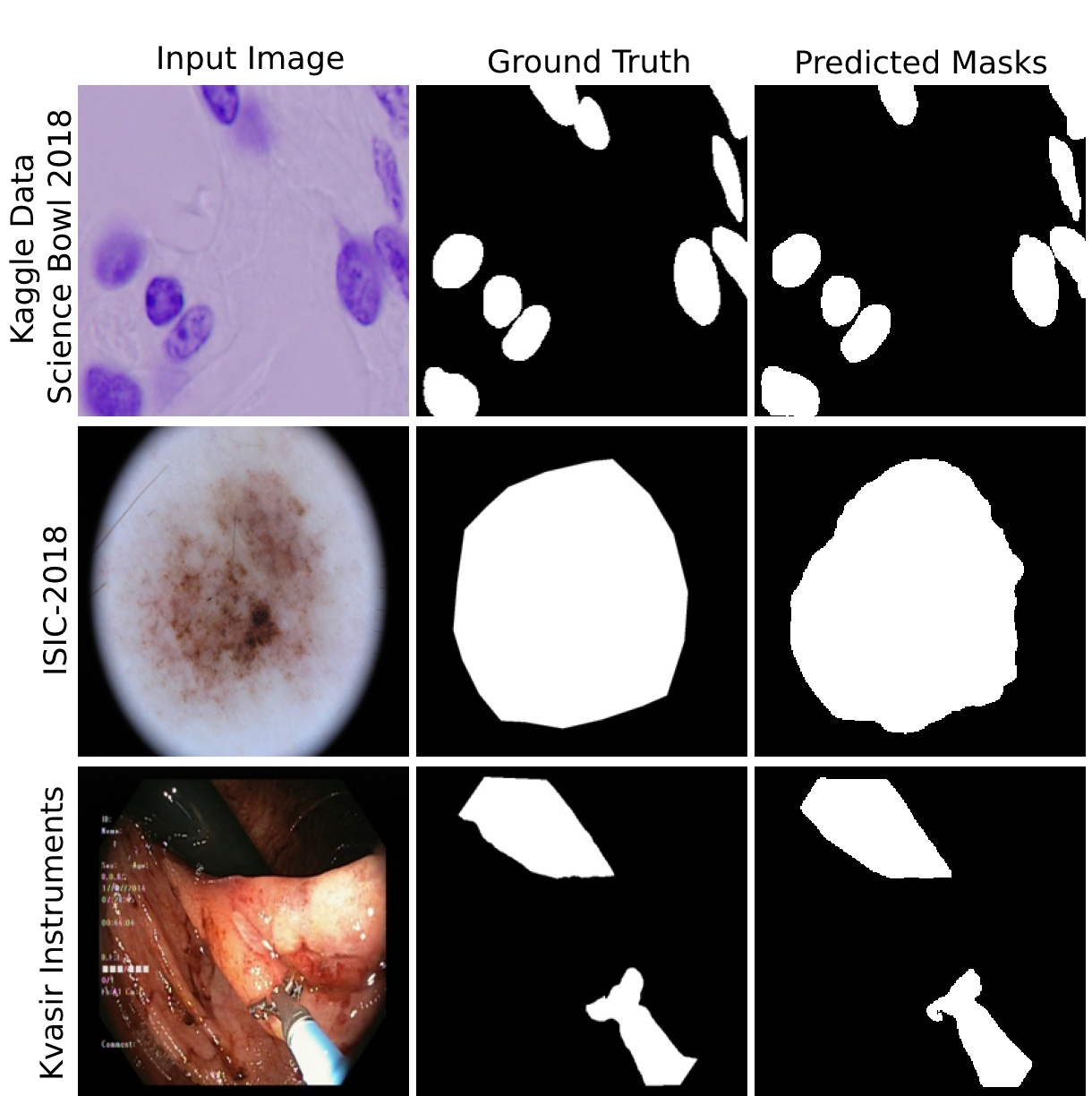}
    \caption{Qualitative results of our proposed PAANet.}
    \label{fig:quali}
    \vspace{-2mm}
\end{figure}

Table~\ref{tab:result1} reports the results obtained by our PAANet, and we can observe that our PAANet achieves a \ac{DSC} of 0.9244 and \ac{mIoU} of 0.8627, outperforming SOTA methods in all the  metrics.  From Table~\ref{tab:result2}, we can observe that PAANet reports a \ac{DSC} of 0.8912 and recall of 0.9019, outperforming the second best MSRF-Net by 0.88\% in terms of \ac{DSC} and 1.26\% in terms of recall.  In Table~\ref{tab:result3}, it can be noted that PAANet attains an improvement of 1.16\% in \ac{DSC} and 1.70\% in \ac{mIoU} over the second best performing MSRF-Net. The qualitative results of our approach can be observed in Figure~\ref{fig:quali}. From the qualitative results, it can be observed that the output segmentation masks produced by PAANet is similar to the ground truth.  Overall, the progressive alternating attention mechanism used to control and limit the features contributed by convolutional layers in residual dense blocks enhances the segmentation ability of our PAANet. The consecutive GAMs which are progressively updated enables the generation of feature maps germane to the region-of-interest and subsequent mining of boundary cues generates detailed and spatially accurate segmentation maps.

\section*{Conclusion}
In this work, we proposed PAANet which employs a novel progressive alternating attention dense block. The PAAD uses an attention mechanism which alternatively applies spatial attention and reverses spatial attention to guide the features contributed by the convolutional layers in the dense block. The attention maps enable layers to focus on the spatial locations pertinent to the target structure. The reverse attention guides the layers to capture the boundary and edge information which further refines the features and allows the generation of finer and spatially precise segmentation maps. Experiments performed on three biomedical image segmentation datasets validated our approach and can be seen as new benchmarks. Our future work will comprise of modifying our PAANet for medical image classification.

\section*{Acknowledgment}
The computations in this paper were performed on equipment provided by the Experimental Infrastructure for Exploration of Exascale Computing (eX³), which is financially supported by the Research Council of Norway under contract 270053.

\bibliographystyle{IEEEtran}
\bibliography{references} 
\end{document}